\documentclass[aps,prd]{revtex4-2}
\usepackage{amsmath, amssymb, amsthm}
\usepackage{physics}
\usepackage{hyperref}
\usepackage{geometry}
\begin{document}

\title{Black Hole Interior Operators and Dilatation Symmetry in Planar Black Branes}
\author{Nirmalya Kajuri}%
 \email{nirmalya@iitmandi.ac.in}
\affiliation{%
School of Physical Sciences, IIT Mandi,\\ Himachal Pradesh 175005, India}%

\begin{abstract}

 Planar AdS black branes have a scaling symmetry that maps a brane solution at one temperature to a solution at another. It is natural to expect that boundary representations of bulk field modes should inherit this symmetry i.e. their correlators should transform covariantly under boundary dilatations. We derive a covariance condition that any boundary representation of interior modes in a planar AdS black brane should satisfy. We then show that Papadodimas-Raju mirror operators satisfy this condition. Thus the Papadodimas-Raju reconstruction of the bulk interior, although state-dependent, inherits the scaling symmetry of planar AdS black holes. 
\end{abstract}
\maketitle
\section{Introduction}
Planar AdS black branes possess a scaling symmetry: the simultaneous rescaling
\begin{equation*}
r \to r/\lambda, \quad t \to \lambda t, \quad \vec{x} \to \lambda \vec{x}
\end{equation*}
leaves the metric invariant up to a rescaling of the horizon radius $r_h \to r_h/\lambda$,
and hence of the Hawking temperature $T \to T/\lambda$~\cite{Gubser:1996de}.
Since this rescaling acts on the boundary coordinates as $(t, \vec{x}) \to (\lambda t, \lambda \vec{x})$,
it corresponds to a dilatation of the boundary CFT that lives on $\mathbb{R}^{d,1}$.
A well-known consequence is that physical observables --- transport coefficients and
quasinormal mode frequencies --- depend only on scaling-covariant dimensionless
ratios~\cite{Horowitz:1999jd, Policastro:2002se, Policastro:2002tn, Son:2007vk,
Policastro:2001yc, Kovtun:2004de, Kovtun:2005ev}.
This symmetry is absent for spherical black holes, whose boundary duals are thermal states that do not enjoy
dilatation invariance.

It is natural to ask whether this symmetry extends to bulk reconstruction.
For exterior reconstruction, the HKLL map~\cite{Dobrev:1998md,Bena:1999jv,Hamilton:2005ju,Hamilton:2006az,Hamilton:2006fh,Kajuri:2020vxf} provides a
state-independent construction of bulk fields outside the horizon, and covariance of
exterior operators under dilatations follows in a relatively straightforward manner (see \cite{Papadodimas:2012aq,Guica:2014dfa,Kajuri:2020bvi,Dey:2024syw} for explicit reconstructions of exterior operators).
The interior is more subtle. Both the mirror operator construction of
Papadodimas and Raju (PR)~\cite{Papadodimas:2013jku, Papadodimas:2013wnh,
Papadodimas:2015jra, Papadodimas:2015xma} and the more recent non-isometric
encoding proposal~\cite{Akers:2022qdl,DeWolfe:2023iuq,Antonini:2023hdh,DeWolfe:2023jfm,Antonini:2024yif,Bueller:2024zvz} are state-dependent, and it is
therefore a non-trivial question whether interior operators can be consistently defined
so as to transform covariantly under boundary dilatations at all.

In this paper we address this question. We first show that demanding covariance of
boundary correlators under dilatations constrains interior operators to satisfy a
precise transformation law, either in the strong form
\begin{equation*}
U_\lambda \widetilde{O}^{(\Psi_T)}_{\omega,\vec{k}}\, U_\lambda^\dagger
= \lambda^{\Delta - d - 1}\, \widetilde{O}^{(\Psi_{T/\lambda})}_{\omega/\lambda,\,\vec{k}/\lambda}
\quad \text{on } \mathcal{H}_{\rm code}(T/\lambda),
\end{equation*}
or a weaker projected version thereof. We then verify that the PR mirror operators
satisfy the stronger condition. This shows that the PR interior reconstruction is fully
consistent with the scaling symmetry of planar AdS black branes. The non-isometric proposal does not provide a concrete prescription that can be checked against our criterion. We argue that in this case, there should be an additional scaling condition purely in the bulk. 

\section{Dilatation Symmetry and Its Action on the Code Subspace}
In this section, we spell out the action of dilatations at three levels: on Fourier modes of primary operators, on black brane microstates, and on the small algebra and its associated code subspace.

The dual to a $d+2$-dimensional planar AdS black brane is a CFT on $\mathbb{R}^{d,1}$. This CFT has a dilatation operator $D$ and corresponding unitary
$U_\lambda = e^{i\alpha D}$ with $\lambda = e^\alpha$. Its action on the boundary coordinates is given by:
\begin{equation}
    (t, \vec{x}) \;\longmapsto\; (\lambda t, \lambda \vec{x}),
\end{equation}
and on a primary operator of dimension $\Delta$:
\begin{equation}
    U_\lambda\, \mathcal{O}(t, \vec{x})\, U_\lambda^\dagger
    = \lambda^\Delta\, \mathcal{O}(\lambda t, \lambda \vec{x}).\label{prim}
\end{equation}

The Hamiltonian $H$ satisfies $[D, H] = -iH$, so
\begin{equation}
    U_\lambda\, H\, U_\lambda^\dagger = \lambda H. \label{ham}
\end{equation}

\paragraph{Action on Fourier modes:}
The  Fourier mode of a primary of dimension $\Delta$ is given by:
$$\mathcal{O}_{\omega, \vec{k}} = \int\, d^d x \,dt\, e^{i\omega t-i\vec{k}\cdot\vec{x}}\,
\mathcal{O}(t, \vec{x})$$

Following \eqref{prim}, this transforms as
\begin{equation}
    U_\lambda\, \mathcal{O}_{\omega,\vec{k}}\, U_\lambda^\dagger
    = \lambda^{\Delta-d-1}\, \mathcal{O}_{\omega/\lambda, \vec{k}/\lambda} 
    \label{eq:mode-transform}
\end{equation}

\paragraph{Action on Black Hole:}

First, let's recall the scaling symmetry on the bulk side.
The metric of a  planar AdS$_{d+2}$ Schwarzschild black brane can be written as:

\begin{equation}
ds^2
=
-r^2 f(r)\,dt^2
+\frac{dr^2}{r^2 f(r)}
+r^2 d\vec x^{\,2},
\qquad
f(r)=1-\left(\frac{r_h}{r}\right)^{d+1} .
\end{equation}
The metric is invariant under the scaling transformation
\begin{equation}
r \;\to\; r/\lambda ,
\qquad
t \;\to\; \lambda t,
\qquad
\vec x \;\to\; \lambda \vec x,
\qquad
r_h \;\to\;  r_h/\lambda . \label{slamb}
\end{equation}

So the metric is mapped to the same functional form, but with horizon radius
\begin{equation}
r_h \to r_h/\lambda  ,
\end{equation}
As the Hawking temperature is proportional to the horizon radius,
\begin{equation}
T=\frac{(d+1)\, r_h}{4\pi}.
\end{equation}
this implies a change of temperature:
\begin{equation}
T \to T/\lambda  .
\end{equation}
\paragraph{Action on black-brane equilibrium states.}
To discuss normalizable microstates and entropy, we temporarily
introduce an infrared regulator by placing the CFT in a spatial box
of linear size $L$.  Let $\mathcal H_L$ denote the corresponding
Hilbert space.  A black-brane equilibrium microstate at temperature
$T$ is taken from a narrow band $\mathcal B_{T,L}$ centred on the
thermodynamic energy $E_0(T,L)$:
\begin{equation}
|\Psi_{T,L}\rangle
=
\sum_{E\in\mathcal B_{T,L}}\sum_a
C_{E,a}\,|E,a;L\rangle,
\qquad
\sum_{E,a}|C_{E,a}|^2=1.
\end{equation}
Here $a$ labels the degeneracy at fixed energy.  We do not assume
that the individual coefficients $C_{E,a}$ have a Boltzmann form.

The property required below is that $|\Psi_{T,L}\rangle$ is an
equilibrium state for the chosen small set of observables.  Namely,
for every normalized $A\in\mathcal A_{\rm small}(T,L)$,
\begin{equation}
\langle\Psi_{T,L}|A|\Psi_{T,L}\rangle
=
\langle A\rangle_{\beta,L}
+\delta_A,
\qquad
\delta_A
=
O(1/N)+O\!\left(e^{-S(T,L)/2}\right),
\end{equation}
where $\langle A\rangle_{\beta,L}$ is the thermal expectation value
at $\beta=1/T$.  The estimate is understood for observables whose
thermal correlators have been normalized to be $O(1)$; it expresses
the working accuracy of bulk reconstruction and the typical
state-to-state fluctuations within the band.

On the infinite plane the dilation is implemented by $U_\lambda$.
With the finite-volume regulator it should be regarded as a unitary
identification
\begin{equation}
U_{\lambda,L}:\mathcal H_L\longrightarrow\mathcal H_{\lambda L},
\qquad
U_{\lambda,L}H_LU_{\lambda,L}^\dagger
=
\lambda H_{\lambda L}.
\end{equation}
Thus an energy eigenstate is carried to an eigenstate of energy
$E/\lambda$ in $\mathcal H_{\lambda L}$.  Up to a unitary change of
basis within degenerate energy eigenspaces,
\begin{equation}
U_{\lambda,L}|E,a;L\rangle
=
\sum_b V_{ba}(E)
|E/\lambda,b;\lambda L\rangle .
\end{equation}

Define
\begin{equation}
|\Psi_{T/\lambda,\lambda L}\rangle
:=
U_{\lambda,L}|\Psi_{T,L}\rangle .
\end{equation}
For
\begin{equation}
A'
=
U_{\lambda,L} A U_{\lambda,L}^\dagger
\in
\mathcal A_{\rm small}(T/\lambda,\lambda L),
\end{equation}
unitarity gives
\begin{equation}
\langle\Psi_{T/\lambda,\lambda L}|A'|
\Psi_{T/\lambda,\lambda L}\rangle
=
\langle\Psi_{T,L}|A|\Psi_{T,L}\rangle .
\end{equation}
Thermal correlators transform in the same way, with
$\beta'=\lambda\beta$.  Hence the deviations $\delta_A$ are
transported with the same parametric suppression: the dilation
introduces no new error.  It therefore maps the class of equilibrium
microstates at $(T,L)$ onto the class at $(T/\lambda,\lambda L)$.
We suppress the regulator label $L$ below.

\paragraph{Action on code subspace/small algebra:}
Given a black brane microstate $|\Psi_T\rangle$ at temperature $T$,
the small algebra is defined as \begin{equation}
\begin{split}
\mathcal A_{\rm small}
(T;\omega_*,k_*,K,E_*)
:=
\operatorname{span}\Bigl\{
&\mathbf 1,\,
\mathcal O_{\omega_1,\vec k_1}\cdots
\mathcal O_{\omega_n,\vec k_n}:\\
&n\leq K,\quad
|\omega_i|<\omega_*,\quad
|\vec k_i|<k_*,\quad
\sum_i|\omega_i|\leq E_*
\Bigr\}.
\end{split}
\end{equation}
We take this set to be closed under Hermitian conjugation.  The
cutoffs $K$ and $E_*$ are chosen so that its dimension is much
smaller than that of the black-brane Hilbert space. The associated state-dependent code subspace is
\begin{equation}
\mathcal H_{\rm code}(T)
=
\mathcal A_{\rm small}(T)|\Psi_T\rangle .
\end{equation}
The exterior modes used to generate this space are state-independent,
whereas $\mathcal H_{\rm code}(T)$ itself depends on the reference
state $|\Psi_T\rangle$.

From \eqref{eq:mode-transform}, the dilatation $U_\lambda$ maps
$\mathcal{O}_{\omega,\vec{k}} \mapsto \lambda^{\Delta-d-1}\mathcal{O}_{\omega/\lambda,\vec{k}/\lambda}$.
The image of the algebra $\mathcal{A}(T, \omega_*, k_*)$ under conjugation by
$U_\lambda$ is therefore
\begin{equation}
    U_\lambda
\mathcal A_{\rm small}(T;\omega_*,k_*,K,E_*)
U_\lambda^\dagger
=
\mathcal A_{\rm small}
\left(
T/\lambda;
\omega_*/\lambda,
k_*/\lambda,
K,
E_*/\lambda
\right).\label{eq:algmap}
\end{equation}

The ratio 
\begin{equation}
    q := \left(\frac{\omega}{T},\frac{\vec{k}}{T}\right)
\end{equation}
remains constant under scaling. Thus each $q$ defines an orbit of dilatation.

Now that we have established how exterior operators transform under dilatations, we proceed to ask what the analogous transformation law must be for operators representing fields in the black brane interior.

\section{Correlator Covariance For Interior Operators}

For correlators of operators representing the exterior, covariance
under dilatations is immediate from \eqref{eq:mode-transform}. Operators inside
the horizon, however, do not admit a state-independent prescription. 
We
 assume only that a boundary representation $\widetilde{\mathcal{O}}^{(\Psi_T)}_{\omega,\vec k}$ of interior modes
exists: a family of CFT operators
, possibly
state-dependent, whose correlators within the code subspace
$\mathcal{H}_{\rm code}(T)$ reproduce the bulk EFT correlators of the
corresponding interior mode on the black-brane background at
temperature $T$, to the standard $O(1/N)$ accuracy of bulk
reconstruction. This defining property fixes every matrix element of
$\widetilde{\mathcal{O}}^{(\Psi_T)}_{\omega,\vec k}$ between
code-subspace states, and hence fixes the projected operator
$P_{\rm code}(T)\,\widetilde{\mathcal{O}}^{(\Psi_T)}_{\omega,\vec k}\,
P_{\rm code}(T)$ uniquely for any admissible construction, where
$P_{\rm code}(T)$ denotes the orthogonal projector onto
$\mathcal{H}_{\rm code}(T)$. In what follows we denote exterior
representations by $\mathcal{O}$ and interior ones by
$\widetilde{\mathcal{O}}$ \footnote{Note that statements concerning
reconstructed operators below are understood as statements within
this code subspace.}.
 
Covariance of correlators with interior insertions is then not an
independent assumption; it follows from the bulk scaling map. The
transformation \eqref{slamb} maps the black-brane solution at
temperature $T$ exactly onto the solution at $T/\lambda$, acting on the
\emph{entire} spacetime, interior included. Bulk EFT correlation
functions on the two backgrounds are therefore identified by this map,
with an interior mode of boundary quantum numbers $(\omega,\vec k)$ on
the $T$ background carried to the mode
$(\omega/\lambda,\vec k/\lambda)$ on the $T/\lambda$ background. The labels and normalization of the interior partner mode are fixed
by the same boundary normalization as the corresponding exterior
mode.  In particular, writing
\begin{equation}
a:=\Delta-d-1,
\end{equation}
the bulk scaling map carries the normalized partner mode according to
\begin{equation}
\widetilde{\mathcal O}^{\rm bulk,(T)}_{\omega,\vec k}
\longmapsto
\lambda^a\,
\widetilde{\mathcal O}^{\rm bulk,(T/\lambda)}
_{\omega/\lambda,\vec k/\lambda}.
\end{equation}
Note that this is a statement about the bulk mode whose boundary representation
is being reconstructed, rather than an assumption about that boundary
representation. The
associated weight is the same $\lambda^{\Delta-d-1}$ as for exterior
modes in \eqref{eq:mode-transform}: in the two-sided description an interior
free-field mode is a linear combination of left and right boundary
modes whose coefficients depend only on the dimensionless ratios
$(\omega/T,\vec k/T)$, each boundary mode carrying weight
$\lambda^{\Delta-d-1}$ under the diagonal dilatation, and a single-sided
representation must reproduce the same correlators. For example, in the two-sided description the interior field has the
schematic mode expansion
\begin{equation}
\phi_{\rm int}^{(T)}
\sim
\int d\omega\,d^dk\,
\left[
f^{(1)}\!\left(
\frac{\omega}{T},
\frac{\vec k}{T},
\frac{r}{T}
\right)
\mathcal O_{\omega,\vec k}
+
f^{(2)}\!\left(
\frac{\omega}{T},
\frac{\vec k}{T},
\frac{r}{T}
\right)
\widetilde{\mathcal O}_{\omega,\vec k}
+\mathrm{h.c.}
\right].
\end{equation}
The coefficient functions depend only on scaling-invariant arguments,
so the exterior and partner modes carry the same Fourier-mode weight.

Meanwhile,
Section~II established that $U_\lambda$ transports the microstate, the
small algebra, and hence the code subspace covariantly:
$|\Psi_T\rangle \mapsto |\Psi_{T/\lambda}\rangle$,
$\mathcal{A}(T,\omega_*) \mapsto
\mathcal{A}(T/\lambda,\omega_*/\lambda)$, and
$\mathcal{H}_{\rm code}(T) \mapsto \mathcal{H}_{\rm code}(T/\lambda)$.
Combining these statements, the code-subspace correlators that an
admissible representation at $T/\lambda$ must reproduce are precisely
the $U_\lambda$-transforms of those at $T$. For correlators with a
single interior insertion this reads
\begin{equation}
\Big\langle \Psi_{T/\lambda}\Big|\,(A'_L)^\dagger\,
\widetilde{\mathcal{O}}^{(\Psi_{T/\lambda})}_{\omega/\lambda,\vec k/\lambda}\,
A'_R\,\Big|\Psi_{T/\lambda}\Big\rangle
= \lambda^{-(\Delta-d-1)}\,
\Big\langle \Psi_T\Big|\,(A_L)^\dagger\,
\widetilde{\mathcal{O}}^{(\Psi_T)}_{\omega,\vec k}\,
A_R\,\Big|\Psi_T\Big\rangle ,
\qquad
A'_{L,R} := U_\lambda A_{L,R}\,U_\lambda^\dagger ,
\label{eq:corrcov}
\end{equation}
for all $A_L, A_R \in \mathcal{A}(T,\omega_*)$. Equation \eqref{eq:corrcov} is thus a necessary condition on
any boundary representation of interior modes, state-dependent or not.
We now recast it as an operator statement.
Define the difference operator
\begin{equation}
D_{\omega,\vec k}
:=
U_\lambda
\widetilde{\mathcal O}^{(\Psi_T)}_{\omega,\vec k}
U_\lambda^\dagger
-
\lambda^{\Delta-d-1}
\widetilde{\mathcal O}^{(\Psi_{T/\lambda})}
_{\omega/\lambda,\vec k/\lambda}.
\end{equation}
Using
$U_\lambda|\Psi_T\rangle=|\Psi_{T/\lambda}\rangle$,
the correlator condition becomes
\begin{equation}
\langle\Psi_{T/\lambda}|
(A_L')^\dagger
D_{\omega,\vec k}
A_R'
|\Psi_{T/\lambda}\rangle
= 0
\end{equation}
for all $A_L',A_R'$ in the transformed small operator set.

Because the set is closed under Hermitian conjugation, vectors of the
form $A_R'|\Psi_{T/\lambda}\rangle$ span
$\mathcal H_{\rm code}(T/\lambda)$ and the corresponding bras span
its dual.  Therefore
\begin{equation}
\langle\phi|D_{\omega,\vec k}|\chi\rangle
= 0,
\qquad
|\phi\rangle,|\chi\rangle
\in\mathcal H_{\rm code}(T/\lambda),
\end{equation}
or equivalently
\begin{equation}
P_{\rm code}(T/\lambda)
D_{\omega,\vec k}
P_{\rm code}(T/\lambda)
= 0.
\end{equation}
Substituting the definition of $D_{\omega,\vec{k}}$, we find that the interior operator must transform covariantly after projection to the code subspace:
\begin{equation}
\label{eq:projected-cov}
\boxed{P_{\rm code}(T/\lambda)\,
U_\lambda \widetilde{\mathcal O}^{(\Psi_T)}_{\omega,\vec{k}} U_\lambda^\dagger\,
P_{\rm code}(T/\lambda)
=
\lambda^{\Delta-d -1}
P_{\rm code}(T/\lambda)\,
\widetilde{\mathcal O}^{(\Psi_{T/\lambda})}_{\omega/\lambda,\vec{k}/\lambda}\,
P_{\rm code}(T/\lambda).}
\end{equation}

In addition, if both
\[
U_\lambda \widetilde{\mathcal O}^{(\Psi_T)}_{\omega,\vec{k}} U_\lambda^\dagger
\qquad\text{and}\qquad
\widetilde{\mathcal O}^{(\Psi_{T/\lambda})}_{\omega/\lambda,\vec{k}/\lambda}
\]
preserve $\mathcal H_{\rm code}(T/\lambda)$, then
\begin{equation}
\label{eq:strongcov}
\boxed{U_\lambda \widetilde{\mathcal O}^{(\Psi_T)}_{\omega,\vec{k}} U_\lambda^\dagger
=
\lambda^{\Delta-d -1}
\widetilde{\mathcal O}^{(\Psi_{T/\lambda})}_{\omega/\lambda,\vec{k{/\lambda}}}
\qquad
\text{on }\mathcal H_{\rm code}(T/\lambda)}
\end{equation} 

Equation \eqref{eq:projected-cov} is the central constraint of our analysis.
Given the spanning property of the code subspace, it is in fact
equivalent to the correlator condition \eqref{eq:corrcov}: the content
of this section is to convert the covariance of code-subspace
correlators, derived above from the bulk scaling map and the defining
accuracy of bulk reconstruction, into an operator criterion against
which concrete constructions can be tested. Any boundary representation
of interior bulk fields, regardless of the specific construction, must
satisfy \eqref{eq:projected-cov} to $O(1/N)$; since the defining property
fixes the projected operator uniquely, \eqref{eq:projected-cov} states that
this unique projection transforms covariantly. If the interior
operators additionally preserve the code subspace --- as the PR mirrors
do up to edge effects at the cutoff $\omega_*$, which itself scales
covariantly by \eqref{eq:algmap} --- the stronger condition
\eqref{eq:strongcov} holds. In the next section we verify that the PR
construction satisfies \eqref{eq:strongcov} directly, without invoking
\eqref{eq:corrcov} at any point.

For the truncated PR operator set it is useful to distinguish an
inner code subspace from the full cutoff space.  For example, let
$\mathcal H_{\rm code}^{\rm in}(T)$ be generated by monomials of
degree at most $K-1$ and total excitation energy at most
$E_*-\omega_*$.  Acting with one additional exterior or mirror mode
then maps
\begin{equation}
\mathcal H_{\rm code}^{\rm in}(T)
\longrightarrow
\mathcal H_{\rm code}(T).
\end{equation}
Statements that an operator ``preserves the code subspace'' are
understood in this nested sense when a sharp truncation is used.

\section{Action on mirror operators}
In this section, we show that the PR mirror operators satisfy \eqref{eq:strongcov}. We give two proofs. The first uses the explicit definition of the mirror operator directly, while the second uses the Tomita-Takesaki modular theory and shows that the result is a consequence of the algebraic properties of the construction rather than specific details of the state.

The PR definition of mirror operator is as follows. Given the small algebra $\mathcal{A}$ and the state $|\Psi_T\rangle$, the
code subspace/little Hilbert space is $\mathcal{H}_\mathrm{code} = \mathcal{A}|\Psi_T\rangle$.
The mirror operator $\widetilde{\mathcal{O}}_{\omega,\vec{k}}$ is defined by its
action on $\mathcal{H}_\mathrm{code}$: for all $A \in \mathcal{A}$,
\begin{equation}
    \widetilde{\mathcal{O}}_{\omega,\vec{k}}\,(A|\Psi\rangle)
    \;=\; A\, e^{-\beta\omega/2}\, \mathcal{O}_{\omega, \vec{k}}^\dagger\,|\Psi\rangle.
    \label{eq:mirror-def}
\end{equation}

Now we check that mirror operators transform covariantly under dilatations: 

\[
U_\lambda \Big[ A \cdot e^{-\beta \omega/2}\,\mathcal{O}_{\omega, \vec{k}}^\dagger \, |\Psi_T\rangle \Big]
= e^{-\beta \omega/2}\, A' \cdot \lambda^{\Delta-d - 1}\,
\mathcal{O}_{\omega/\lambda, \vec{k}/\lambda}^\dagger \,|\Psi_{T/\lambda}\rangle
\]

where $A'=U_\lambda A U^\dagger_\lambda $ and $|\Psi_{T/\lambda}\rangle =U_\lambda |\Psi_T\rangle .$

So:
\[
\big(U_\lambda \,\widetilde{\mathcal{O}}_{\omega,\vec{k}}\, U_\lambda^\dagger\big)
\big(A'|\Psi_{T/\lambda}\rangle\big)
= A' \cdot \lambda^{\Delta-d - 1}\,
e^{-\beta \omega/2}\,
\mathcal{O}_{\omega/\lambda, \vec{k}/\lambda}^\dagger |\Psi_{T/\lambda}\rangle
\]

Now compare with what the correct mirror operator formula demands at temperature
$T'=T/\lambda$ ($\beta^\prime=\lambda\beta$) and, $\omega' = \omega/\lambda$ and momentum $\vec{k}'=\vec{k}/\lambda$:
\[
\widetilde{\mathcal{O}}^{(\Psi_{T/\lambda})}_{\omega/\lambda,\vec{k}/\lambda}
\big(A'|\Psi_{T/\lambda}\rangle\big)
= A' \cdot e^{-\beta'\omega'/2}\,
\mathcal{O}_{\omega', \vec{k'}}^\dagger |\Psi_{T/\lambda}\rangle
= A' \cdot
e^{-(\beta\lambda) (\omega/(2\lambda))}\,
\mathcal{O}_{\omega/\lambda, \vec{k}/\lambda}^\dagger |\Psi_{T/\lambda}\rangle=A' \cdot \,
e^{-\beta \omega/2}\,
\mathcal{O}_{\omega/\lambda, \vec{k}/\lambda}^\dagger |\Psi_{T/\lambda}\rangle
\]
 
This gives us:
\begin{equation}
U_\lambda \widetilde {\mathcal{O}}_{\omega,\vec{k}} U_\lambda^\dagger
=
\lambda^{\Delta-d-1}\widetilde{ \mathcal{O}}^{(\Psi_{T/\lambda})}_{\omega/\lambda,\vec{k}/\lambda}\qquad
\text{on }\mathcal H_{\rm code}(T/\lambda).\label{mirrort}
\end{equation}
Thus the transformed mirror has the same defining action as the
mirror associated with $|\Psi_{T/\lambda}\rangle$, on the transformed
code subspace.

We now give a complementary algebraic explanation of the covariance
property using Tomita-Takesaki theory The relation between PR mirror operators and Tomita--Takesaki
modular conjugation is detailed in \cite{Papadodimas:2013jku}.  We use this relation
here to give a second proof of covariance.  For an adjoint-closed
algebra $\mathfrak A$ with a cyclic and separating state
$|\Psi\rangle$, the identification
\begin{equation}
\widetilde{\mathcal O}^{(\Psi)}_{\omega,\vec k}
=
J_\Psi\mathcal O_{\omega,\vec k}J_\Psi
\end{equation}
is an exact statement on the Hilbert space generated by
$\mathfrak A|\Psi\rangle$. Here $J_\Psi$ is the modular conjugation associated to the state $ |\Psi\rangle $ and
the algebra $\mathcal{A}$. 

In the PR application, the small operator set is truncated and only
approximately closed.  Accordingly, its modular description, and the
equalities derived from it below, are understood as statements about
the action of the operators within the code subspace, to the working
large-$N$ accuracy and away from cutoff-edge effects.  They do not
specify an extension of the mirror operators to the full CFT Hilbert
space.  The direct proof above does not require these additional
modular-theory assumptions.

The Tomita operator $S_\Psi$ is the antilinear operator defined by
\begin{equation}
S_\Psi\bigl(A|\Psi\rangle\bigr)
=
A^\dagger|\Psi\rangle,
\qquad
A\in\mathfrak A .
\end{equation}
First, we show that the Tomita operator transforms covariantly. We conjugate both sides by $U_\lambda$. On the left-hand side,
\begin{equation}
    U_\lambda S_\Psi U_\lambda^\dagger \bigl(U_\lambda A |\Psi\rangle\bigr)
    = U_\lambda S_\Psi \bigl(A|\Psi\rangle\bigr)
    = U_\lambda A^\dagger |\Psi\rangle
    = (U_\lambda A^\dagger U_\lambda^\dagger)\,|\Psi_{T/\lambda}\rangle
    = (A')^\dagger\,|\Psi_{T/\lambda}\rangle,
\end{equation}
where $A' := U_\lambda A U_\lambda^\dagger \in \mathcal{A}(T/\lambda)$ and we used
$U_\lambda|\Psi_T\rangle = |\Psi_{T/\lambda}\rangle$. Since vectors of the form
$A'|\Psi_{T/\lambda}\rangle$ span $\mathcal{H}_{\rm code}(T/\lambda)$, and since
the right-hand side equals $S_{\Psi_{T/\lambda}}(A'|\Psi_{T/\lambda}\rangle)$,
we conclude
\begin{equation}
    U_\lambda\, S_{\Psi_T}\, U_\lambda^\dagger = S_{\Psi_{T/\lambda}}.
    \label{eq:tomita-cov}
\end{equation}
Next, we show that the modular conjugation transforms covariantly. The polar decomposition of the Tomita operator is $S_\Psi = J_\Psi\,\Delta_\Psi^{1/2}$,
where $J_\Psi$ is anti-unitary and $\Delta_\Psi^{1/2}$ is positive. Since $U_\lambda$
is unitary, conjugation by $U_\lambda$ preserves this decomposition. Applying
$U_\lambda\,(\cdot)\,U_\lambda^\dagger$ to both sides of $S_{\Psi_T} = J_{\Psi_T}\Delta_{\Psi_T}^{1/2}$
and using \eqref{eq:tomita-cov} gives
\begin{equation}
    S_{\Psi_{T/\lambda}}
    = \bigl(U_\lambda J_{\Psi_T} U_\lambda^\dagger\bigr)
      \bigl(U_\lambda \Delta_{\Psi_T}^{1/2} U_\lambda^\dagger\bigr).
\end{equation}
By uniqueness of the polar decomposition, we can identify the two factors separately:
\begin{equation}
    J_{\Psi_{T/\lambda}} = U_\lambda\, J_{\Psi_T}\, U_\lambda^\dagger,
    \qquad
    \Delta_{\Psi_{T/\lambda}}^{1/2} = U_\lambda\, \Delta_{\Psi_T}^{1/2}\, U_\lambda^\dagger.
    \label{eq:modular-cov}
\end{equation}

Finally, using \eqref{eq:mode-transform} and \eqref{eq:modular-cov} we have:
\[
\begin{aligned}
U_\lambda 
\widetilde O^{(\Psi_T)}_{\omega,\vec k}
U_\lambda^\dagger
&=
U_\lambda J_{\Psi_T}
O_{\omega,\vec k}
J_{\Psi_T}U_\lambda^\dagger  \\
&=
J_{\Psi_{T/\lambda}}
\left(
U_\lambda O_{\omega,\vec k}U_\lambda^\dagger
\right)
J_{\Psi_{T/\lambda}}  \\
&=
J_{\Psi_{T/\lambda}}
\left(
\lambda^{\Delta-d-1}
O_{\omega/\lambda,\vec k/\lambda}
\right)
J_{\Psi_{T/\lambda}}  \\
&=
\lambda^{\Delta-d-1}
\widetilde O^{(\Psi_{T/\lambda})}_{\omega/\lambda,\vec k/\lambda}.
\end{aligned}
\]
This agrees with \eqref{mirrort}.
The action of dilatations on the interior reconstruction data can therefore
be summarized as a map
\[
R_\lambda:
\left(
\mathcal A(T,\omega_*,k_*),
|\Psi_T\rangle,
\widetilde{\mathcal A}^{(\Psi_T)}(T,\omega_*,k_*)
\right)
\longmapsto
\left(
\mathcal A\!\left(T/\lambda,\omega_*/\lambda,k_*/\lambda\right),
|\Psi_{T/\lambda}\rangle,
\widetilde{\mathcal A}^{(\Psi_{T/\lambda})}
\!\left(T/\lambda,\omega_*/\lambda,k_*/\lambda\right)
\right).
\]
Here \(\widetilde{\mathcal A}^{(\Psi_T)}\) denotes the mirror algebra
associated with the state \(|\Psi_T\rangle\). The corresponding generators
transform as
\[
O_{\omega,\vec k}
\longmapsto
\lambda^{\Delta-d-1}
O_{\omega/\lambda,\vec k/\lambda},
\qquad
\widetilde O^{(\Psi_T)}_{\omega,\vec k}
\longmapsto
\lambda^{\Delta-d-1}
\widetilde O^{(\Psi_{T/\lambda})}_{\omega/\lambda,\vec k/\lambda}.
\]
Thus the entire reconstruction triple transforms covariantly under the
planar black brane scaling symmetry.

An interesting aside about covariant transformation of operators.
If we take an operator string built from small-algebra operators and mirror operators,
\[
\mathcal X_{\{\omega_i,\vec{k}_i\}} = X_{\omega_1,\vec{k_1}} X_{\omega_2,\vec{k_2}} \cdots X_{\omega_n,\vec{k_n}},
\]
where each \(X_{\omega_i,\vec{k_i}}\) is either a small-algebra mode \(O_{\omega_i,\vec{k_i}}\) or a mirror mode \(\widetilde O_{\omega_i,\vec{k_i}}\) with conformal weight $\Delta_i$.

As discussed prior, it follows that

\[\langle \Psi_{T/\lambda}|
\,\mathcal X_{\{\omega_i/\lambda,\vec{k_i}/\lambda\}}\,
|\Psi_{T/\lambda}\rangle
=
\lambda^{-\sum_i(\Delta_i-d-1)}
\langle \Psi_T|\,\mathcal X_{\{\omega_i,k_i\}}|\Psi_T\rangle .
\]

Defining \(\widehat X_{\omega,\vec k}^{(T)}=\beta^{\Delta-d-1}X_{\omega,\vec k}^{(T)}\) for either an exterior or a mirror mode, the corresponding dimensionless correlators depend only on \(q_i=(\omega_i/T,\vec k_i/T)\) and are invariant along a dilatation orbit.

Clearly the PR reconstruction for planar black branes, though state dependent, inherits the scaling symmetry of AdS black branes. 

\section{Discussion}

In this note we used the bulk scaling map, together with the defining
accuracy of a boundary reconstruction, to establish covariance of
code-subspace correlators containing one interior insertion.  We then
showed that this correlator statement is equivalent to a projected
operator criterion on the boundary representation of the interior
mode.  Accordingly, every admissible representation that reproduces
the scale-related bulk-EFT matrix elements must satisfy the projected
criterion to the same accuracy.

For the PR construction we proved the stronger transformation law on
the code subspace directly from the defining action of the mirror
operators, and independently from the general correlator argument.
Thus the state dependence of the PR reconstruction is compatible with
the planar black-brane scaling symmetry.  

Another case of interest is the non-isometric picture, which addresses the apparent overcounting of interior states by noting that exponentially many distinct bulk configurations are virtually indistinguishable to simple boundary observers. This picture suggests that the encoding of the black hole interior Hilbert space $\mathcal H_{\text{interior}}$ into $\mathcal H_{\text{CFT}}$
is a composition of two maps. One first quotients by a space of null states, which are invisible to simple probes. The quotient Hilbert space $\mathcal H_{\rm{code}}$ is then mapped to $\mathcal H_{\text{CFT}}$ via an approximately isometric embedding. From the boundary point of view, the covariance condition \eqref{eq:projected-cov} should still hold, since it is a necessary condition on any CFT representation of interior operators, regardless of whether the underlying encoding is isometric or non-isometric.
However, there is a further natural condition at the level of the bulk description: the bulk scaling $S_\lambda$ induced by \eqref{slamb} should preserve the null subspace itself. This is because null states are defined by indistinguishability under simple probes. As simple probes transform covariantly, the partition of $\mathcal H_\text{interior}$ into detectable and null sectors should itself be covariant---null states at temperature $T$ should be mapped to null states at temperature $T/\lambda$. Since null states are expected to be high-complexity states, $S_\lambda $  should map high(low) complexity states to high(low) complexity states, with any complexity threshold formulated in terms of scaling-covariant dimensionless variables.

This additional requirement would not be visible directly in the boundary CFT, since the CFT only sees the encoded code subspace and not the null-state quotient itself; rather, it would be a structural property of the bulk scaling map $S_\lambda$ in the non-isometric framework. Tensor network or qubit models of planar black branes would provide a concrete setting in which to test whether the covariance of the null-state partition under bulk scaling can be made precise, and whether complexity thresholds respect the dilatation symmetry.

\begin{acknowledgments}
 I am grateful to Suvrat Raju for helpful comments.
\end{acknowledgments}

\bibliographystyle{unsrt}
\bibliography{main}

\end{document}